\documentclass[
reprint,
showkeys,
amsmath,amssymb,
 aps,
]{revtex4-1}

\usepackage{graphicx}
\usepackage{dcolumn}
\usepackage{bm}
\usepackage{caption}
\usepackage{subcaption}
\usepackage{enumitem}
\newcommand{\etal}{{\it et al.}}
\newcommand{\EMD}{$E_{MD}$}
\newcommand{\NF}{$N_F$}
\newcommand{\NDA}{$N_{DA}$}
\newcommand{\Ed}{$E_d$}
\newcommand{\Smax}{$S^{max}_{SIA}$}

\begin{document}

\title{Cascade morphology transition in bcc metals}

\author{Wahyu Setyawan$^1$}
\email{Corresponding author, wahyu.setyawan@pnnl.gov}
\author{Aaron P. Selby$^2$}
\author{Niklas Juslin$^2$}
\author{Roger E. Stoller$^3$}
\author{Brian D. Wirth$^{2}$}
\author{Richard J. Kurtz$^1$}
\affiliation{
$^1$Pacific Northwest National Laboratory, Richland, WA 99354, USA\\
$^2$Department of Nuclear Engineering, University of Tennessee, Knoxville, TN 37996, USA\\
$^3$Oak Ridge National Laboratory, Oak Ridge, TN 37831, USA
}

\date{\today}

\begin{abstract}
Energetic atom collisions in solids induce shockwaves with complex morphologies. In this paper, we establish the existence of a morphological transition in such cascades. The order parameter of the morphology is defined as the exponent, $b$, in the defect production curve as a function of cascade energy ($N_F \sim E_{MD}^b$). Response of different bcc metals can be compared in a consistent energy domain when the energy is normalized by the transition energy, $\mu$, between the high- and the low-energy regime. Using Cr, Fe, Mo and W data, an empirical formula of $\mu$ as a function of displacement threshold energy, $E_d$, is presented for bcc metals.
\end{abstract}

\maketitle

\section{Introduction}
An energetic particle entering a solid metal will initiate a collision cascade upon scattering with an atomic nucleus, and this process often induces complex shockwaves. Recent experimental observations suggest that small dislocation loops may be formed within individual displacement cascades during in-situ TEM with 150-keV self-ion irradiation of W \cite{YiPhilMag13}, which stimulates fundamental questions on how the formation of these extended defects and other defect distributions are related to the underlying  cascade morphology. Furthermore, it is also an interesting question to determine whether morphological transition exists as a function of energy. Computer advancements have enabled molecular dynamics (MD) simulations involving millions of atoms and visualization of very large MD data sets. By mapping the collision paths, Calder \etal\ were able to correlate the distribution of defect clusters in space to the shape of the collision regions \cite{CalderPhilMag10}. They concluded that large self-interstitial-atom (SIA) clusters form due to the intersection of high atomic density regions associated with two supersonic shock waves. When this occurs, the excess atoms from the high-density region of one shock front are deposited into the low-density region associated with the second shock wave.

In the present work, we explore and classify various cascade morphologies and their implications on defect creation by simulating displacement cascades in Cr, Fe, Mo and W as a function of cascade energy. We show the existence of a morphological transition and determine the transition energy in these body-centered cubic (bcc) metals. In addition, a scaling relationship for the transition energy in these metals is explored.

\section{Methods}
The MD simulations are performed using the LAMMPS code \cite{LAMMPS}. The forces are derived mostly from Finnis-Sinclair potentials \cite{FinnisPhilMag84} in which the repulsive parts have been extended for cascade simulations; Fe by Calder \etal\ \cite{CalderJNM93}, Mo by Salonen \etal\ \cite{SalonenJPCM03, AcklandPot} and W by Juslin \etal\ \cite{JuslinHeWpot, AcklandPot}. For Cr, the embedded-atom-method potential by Bonny \etal\ is taken \cite{BonnyPhilMag11} with short-range extension done by Juslin for this work.

Before cascade simulation, the atoms are thermalized at 300 K and zero pressure with a Nos\'{e}-Hoover thermostat to obtain a proper distribution of positions and velocities.
The cascade is initiated by giving a random primary-knock-on-atom (PKA) near the center of the cell an initial velocity with a random direction. In keeping with our comparison to previous work \cite{SelbyJNM13, StollerBook12, StollerJNM97, SetyawanJNM14PartI}, the simulations did not treat electronic stopping. To first order, neglecting electronic losses amounts to an energy shift which is related to the difference between the true PKA energy and the MD cascade energy; the latter of which is essentially equivalent to the damage energy in the Norgett-Robinson-Torrens (NRT) displacement model \cite{StollerBook12, NRT}. Based on the results of more detailed cascade simulation models which include electronic stopping \cite{ZarkadoulaJPCM14, DaraszewiczThesis}, there should be no significant influence on our results except that the phenomena may occur at slightly higher energies.

The PKA initial kinetic energy is denoted by \EMD. An adaptive time step is used with a maximum displacement of 0.005 \AA\ per step. The cascades in Cr and W have been simulated in an NVE ensemble for all atoms except the border atoms. The border atoms are those within one lattice unit from the box edges. A Nos\'{e}-Hoover thermostat at 300 K with 50-fs time constant is applied to the border atoms to extract heat from the system. With this setting, the temperature of the system is $<$ 400 K at the end of the simulation. Such an increase of the temperature has negligible effect on the results presented here. Therefore, the choice of the time constant is appropriate. The simulations are stopped when the number of surviving Frenkel pairs, \NF, does not change for more than 10 ps. Note that for all other metals, unless otherwise specified, the cascade simulations were performed with this default procedure. Some of the cascade data for W are taken from our earlier study \cite{SetyawanJNM14PartI}, in which NVE was used for the first 10 ps, then NVT was applied to all atoms, and the simulation was followed up to ~ 55 ps. We have verified that the two simulation settings yield the same results within the statistical error. For Fe and Mo, the simulations were performed at 100 K and 300 K, respectively, as described in Refs. \cite{StollerBook12, SelbyJNM13}. Sufficiently large simulation cells are used to ensure no interaction between displaced atoms across the periodic boundaries. An SIA or a vacancy is determined from the occupancy of the Wigner-Seitz cells of a reference lattice.

For our analysis, a consistent determination of displacement threshold energy for each potential, \Ed, is essential. Therefore, all values of \Ed\ in this study are calculated using the same method, as described here. Periodic boundaries are applied along all axes. An orthorhombic box of 20$\times$18$\times$16 supercell is used to avoid the self-interaction of the SIA moving along the $<$111$>$ directions. The box contains 11,520 atoms. In test runs using 10 PKAs in each direction along [100], [110], [111], and [136], a smaller box size of 18$\times$16$\times$14 is used and the results are the same within the standard error as with the 20$\times$18$\times$16, giving a confidence that the 20$\times$18$\times$16 box is sufficient. We have also visually verified that replacement events are contained within the box when an SIA is created.

\begin{figure}[hbp]
\centering
\includegraphics[width=0.45\textwidth]{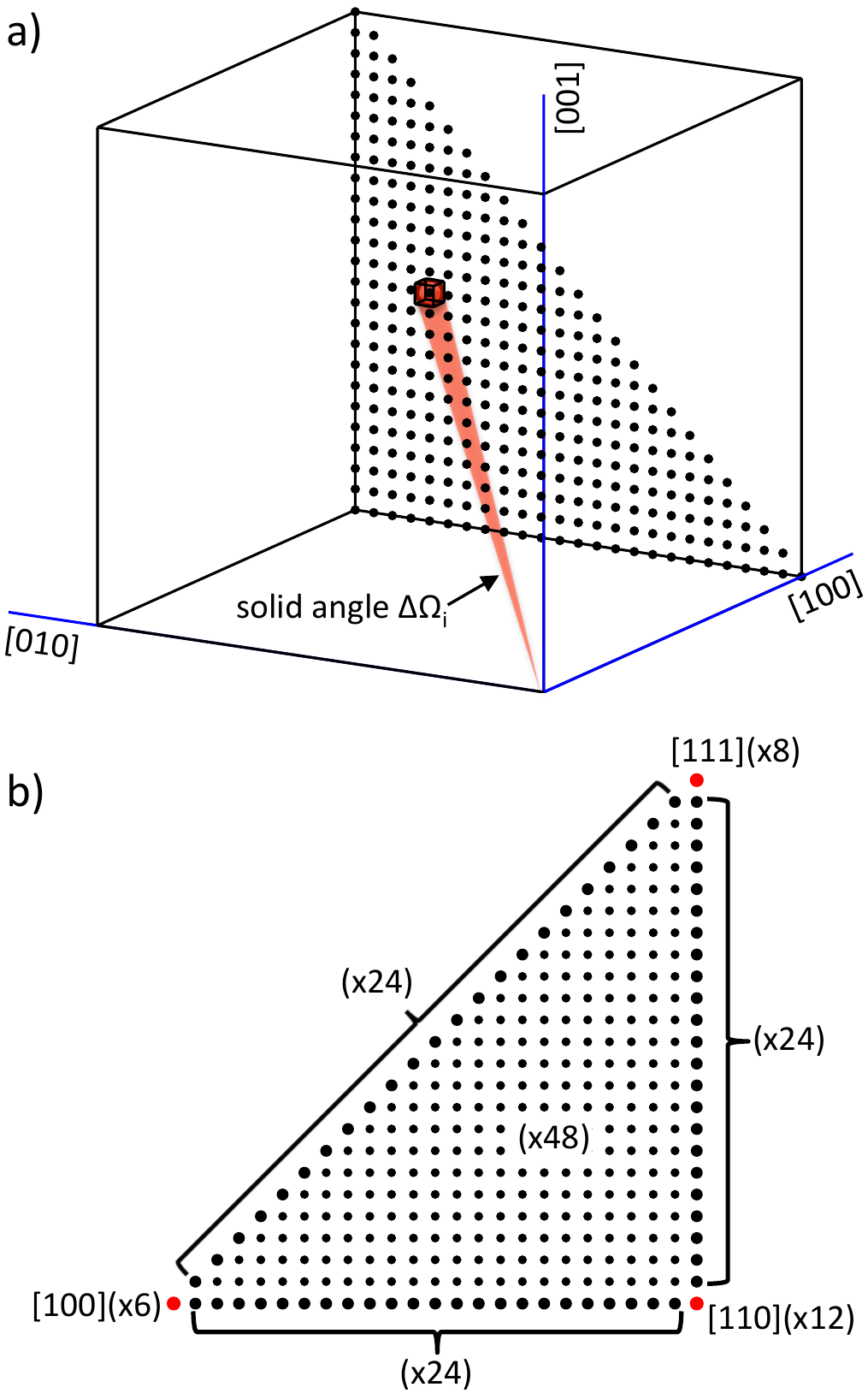}
\caption{a) The irreducible part of the crystal directions and the grid points for the PKA directions. The grids are constructed with a Miller index increment of 1/24 resulting in a total of 325 points. b) The number of equivalent directions of each point is shown as the number following the x character given in parentheses.
\label{fig_grid}}
\end{figure}

For threshold energy simulation, the system is equilibrated at 10 K and zero pressure for 20 ps before a displacement is initiated. Then, low-energy displacements are simulated using the same setting as described previously. In this case, the border atoms are equilibrated at 10 K to extract heat. From visualization, it is evident that 3 ps is appropriate for observing defect creation. Therefore, the simulation is followed up to 3 ps. In the test runs, we have also verified that simulations up to 5 ps give the same results.
Since the high-energy cascades that are analyzed in this study involve collisions along random directions, the appropriate \Ed\ is the average value over all directions \cite{NordlundNIMB06}. The average value is defined as
\begin{equation}
E_d = \frac{1}{4\pi}\int_0^{4\pi} E_d(\Omega)d\Omega,
\label{eq_Ed}
\end{equation}
where $\Omega$ is the direction in space and $d\Omega$ is the solid angle. Due to symmetry, only $E_d(\Omega)$ that are within the irreducible crystal directions (ICD) are needed. The ICD and the PKA direction grid used in this study are shown in Figure \ref{fig_grid}a. 

If ICD are employed, it is necessary to account for the different multiplicity of each direction to correctly calculate the average. The multiplicity represents the number of equivalent directions of a grid point. In Figure \ref{fig_grid}b, the multiplicity of grid point-$i$, denoted as $P_i$, is shown as the number following the x character given in parentheses. The grids are constructed with a Miller index increment of 1/24 resulting in a total of 325 points. Note that these points are not located on the surface of a sphere centered at the origin. Therefore, an appropriate weight due to the different solid angle of each point needs to be taken into account. The solid angle of point-$i$, denoted as $\Delta\Omega_i$, is calculated as the solid angle enclosing a small sphere with a diameter $D$ = 1/24 (i.e. the diameter is the same as the side length of the grid cube) centered at grid point-$i$, as illustrated in Figure \ref{fig_grid}a. Finally, Equation \ref{eq_Ed} becomes
\begin{equation}
E_d = \sum_i E_{d, i}P_i \Delta\Omega_i \Big/ \sum_i P_i \Delta\Omega_i,
\label{eq_Ed2}
\end{equation}
\begin{equation}
\Delta\Omega_i(h, k, l) = 2\pi \left( 1-\frac{\sqrt{h^2+k^2+l^2}}{\sqrt{h^2+k^2+l^2+D^2/4}} \right),
\end{equation}
where $h$, $k$, $l$ are the Miller indices. For each direction, $E_{d, i}$ is averaged from 10 simulation runs, each using a different PKA chosen from near the box center. To search for the threshold energy, the PKA energy is incremented by 5 eV until a defect is detected, then decremented by 1 eV until no defects are detected. The minimum energy where a defect is detected is taken as the threshold for this specific run.

\section{Results and Discussion}
\begin{figure}[hbp]
\centering
\includegraphics[width=0.48\textwidth]{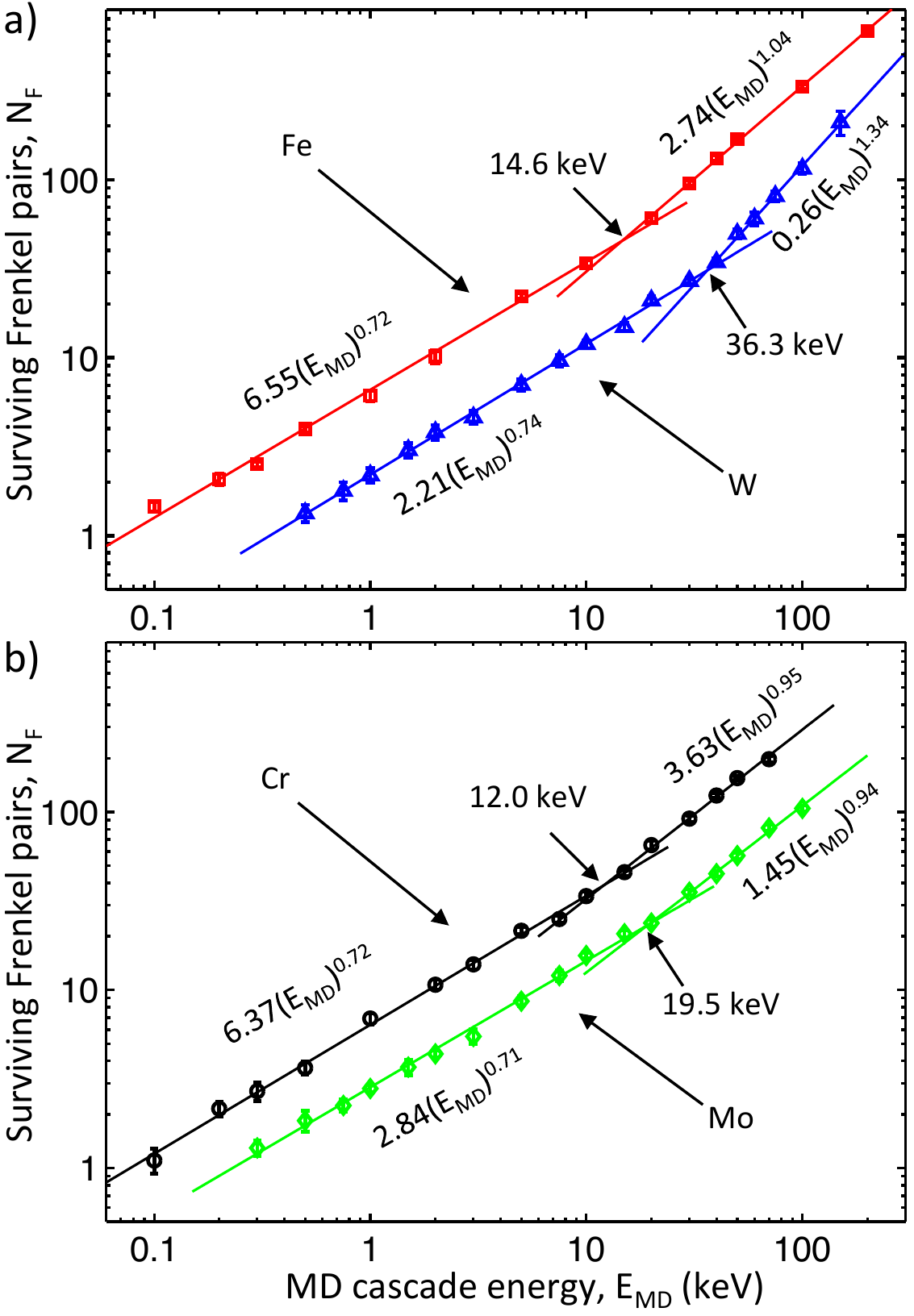}
\caption{(Color online) Plots of \NF\ vs. \EMD\ with standard error. The fit lines reveal two energy regimes with a transition energy $\mu$.
\label{fig_fit}}
\end{figure}

Figure \ref{fig_fit} shows the plot of \NF\ vs. \EMD\ for each metal on a log-log scale. The data points exhibit remarkably clear linear regimes on this scale ($N_F$$\sim$$E_{MD}^b$). It is evident from the W and Fe curves that two linear regions exist, characterized by different slopes, $b$. It is logical then to expect similar behavior for Cr and Mo, which is observed albeit with a less pronounced transition. To consistently determine the characteristic slopes, the data are fit as follows. Only data points where $N_F > 1$ are included in the fit. This is motivated by the fact that in the energy regime where $N_F < 1$, the probability of creating a Frenkel pair is normally either 1 or 0. In such a regime, cascade-like behavior does not occur. A linear regression is performed to fit the data with two lines. The determination of which line segment the data points should be assigned is based on a minimization of the total norm of the residuals from the two fit lines. The value of \EMD\ at the intersection (inflection) of the fit lines is defined as the transition energy, $\mu$. We note that the slopes determined in this way are slightly different from those obtained by the original authors for Fe \cite{StollerBook12} and Mo \cite{SelbyJNM13}.

\begin{figure}[htb]
\centering
\includegraphics[width=0.48\textwidth]{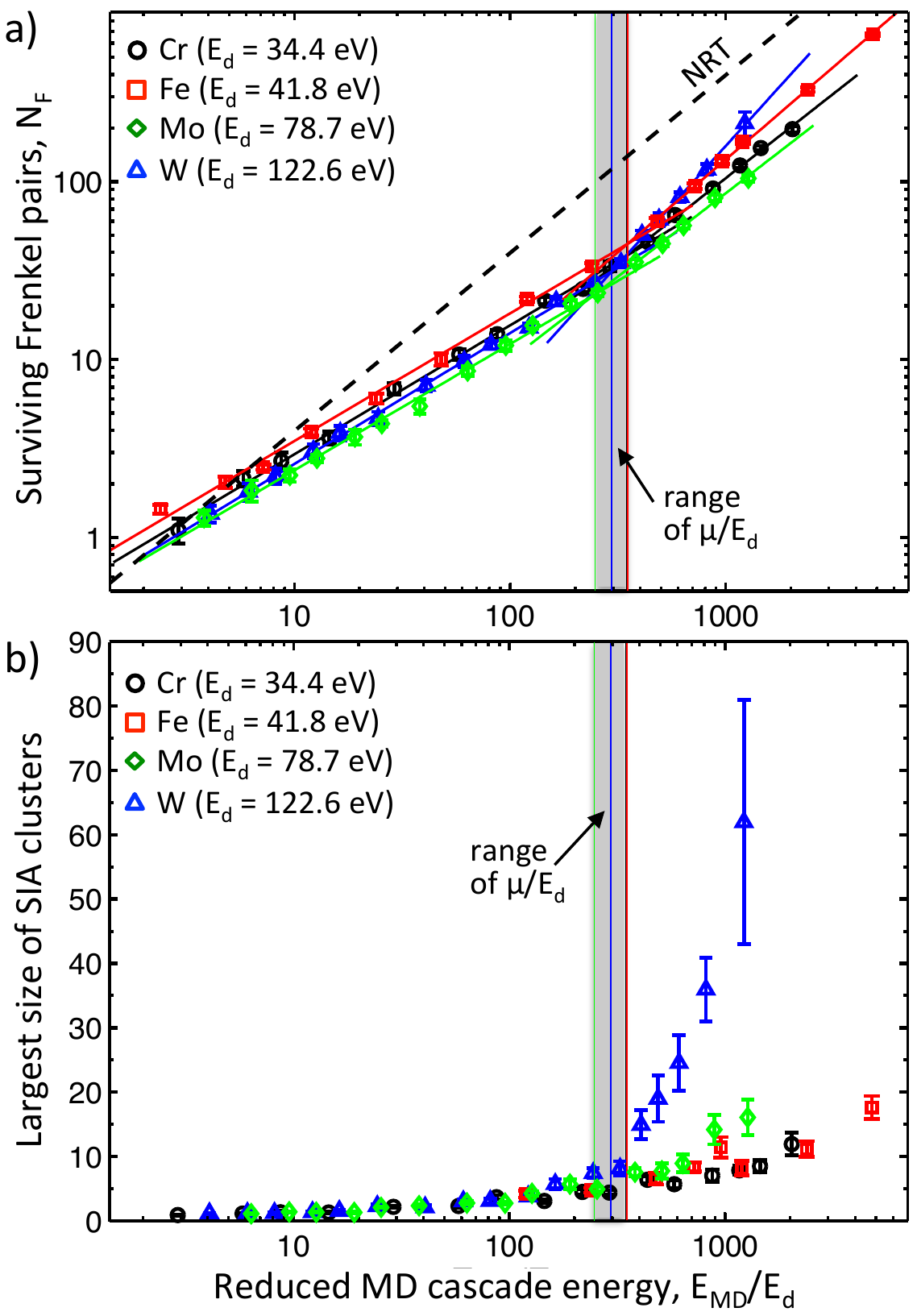}
\caption{(Color online) a) Plots of \NF\ vs. \EMD/\Ed\ with standard error. The range of the transition energy, $\mu$, is shown as a vertical gray bar. b) Average size of the largest SIA clusters, revealing the formation of large SIA clusters in tungsten after the transition energy.
\label{fig_reduced}}
\end{figure}

Due to the different displacement threshold energies of each metal, perhaps a better way to compare different materials is to use a reduced energy $E^*$ = \EMD/\Ed. Table I summarizes the values of \Ed\ calculated with Equation \ref{eq_Ed2}. Figure \ref{fig_reduced}a shows the plots of \NF\ vs. $E^*$. The standard NRT model, \NF\ = 0.4 $E^*$, is included for reference \cite{NRT}. The deviation of MD results from the NRT model is evident. The deviation arises from the fact that the efficiency of in-cascade SIA-vacancy recombination is energy dependent. In the low-energy regime ($E_{MD} < \mu$), in-cascade recombination increases with energy. This leads to sub-linear behavior with a characteristic slope of 0.72, 0.72, 0.71 and 0.74 for Cr, Fe, Mo and W, respectively. Interestingly, the slopes in this low-energy region appear independent of \Ed. In fact, the values are remarkably constant. On the other hand, in the high-energy regime ($E_{MD} > \mu$) the slopes are 0.95, 1.04, 0.94 and 1.34 for Cr, Fe, Mo and W, respectively. While the values for Cr, Fe and Mo are also very similar in this regime, the slope of W is significantly higher. In Figure \ref{fig_reduced}, a vertical gray bar marks the range of $\mu$. When expressed in the reduced unit, $\mu^* = \mu/E_d$, the transition energy falls in a narrow range of 247-350. The narrow range of $\mu^*$ indicates strong self-consistency in the energy fits, and the appropriateness of normalizing defect production in these bcc metals with \Ed. Scaling of $\mu$ as a function of \Ed\ is presented later in this paper.

\begin{table}
\caption{Displacement threshold energy, \Ed, calculated using Equation \ref{eq_Ed2} as described in the text, and using normal or variable atomic masses, as described later in the text.}
\begin{tabular}{lcl}
\hline
Interatomic Potential & \Ed\ (eV) & Note \\
\hline
Cr & 34.4 $\pm$ 1.5 & As is.\\
Fe & 41.8 $\pm$ 1.6 & As is.\\
Mo & 78.7 $\pm$ 2.9 & As is.\\
W & 122.6 $\pm$ 4.4 & As is.\\
\hline
Cr & 33.8 $\pm$ 1.4 & With W mass.\\
Cr & 18.9 $\pm$ 1.1 & With W PKAmass.\\
W & 123.6 $\pm$ 4.4 & With Cr mass.\\
W & 170.9 $\pm$ 6.7 & With Cr PKAmass.\\
W-bj\"{o}rkas & 98.0 $\pm$ 3.7 & As is.\\
\hline
\end{tabular}
\label{table_Ed}
\end{table}

Identifying the underlying physical origin of the change in slope of defect production is non-trivial due to the complex nature of the cascade process. This has previously been associated qualitatively with the onset of sub-cascade formation \cite{StollerBook12, HeinischJNM91}. Here we seek a more detailed explanation based on the cascade damage production process. The size distribution of the surviving defect clusters provides a hint. Interstitial clusters are identified with a cutoff of the third nearest-neighbor distance (NN3) \cite{SetyawanJNM14PartI}. Varying the cutoff from NN1 to NN3 does not alter the qualitative results presented here \cite{SetyawanJNM14PartI}. Figure 3b shows the average size (over all simulations) of the largest SIA clusters, \Smax. A striking correlation can be seen between $\mu^*$ and the onset of the rapid increase in the \Smax\ curves. For $E_{MD} < \mu$ the \Smax\ for all metals is similar, consistent with the similar slopes of defect production in this regime. However, for $E_{MD} > \mu$, the \Smax\ increases considerably. Chromium, Fe and Mo all display a similar increase in \Smax\ that is consistent with the similarity in the slope of defect production in this regime ($b \approx 1.0$), whereas W shows an even more significant increase in \Smax. From Figures \ref{fig_reduced}a-b, it is evident that the slope change of the defect production curve is correlated to the propensity for SIA clustering and the formation of large SIA clusters. This correlation suggests three cascade morphologies as a function of energy, namely i) $E_{MD} < \mu$ ($b < 1$), ii) $E_{MD} > \mu$ of Cr, Fe and Mo ($b \approx 1$), and iii) $E_{MD} > \mu$ of W ($b > 1$).

\begin{figure*}[htbp]
\centering
\includegraphics[width=0.98\textwidth]{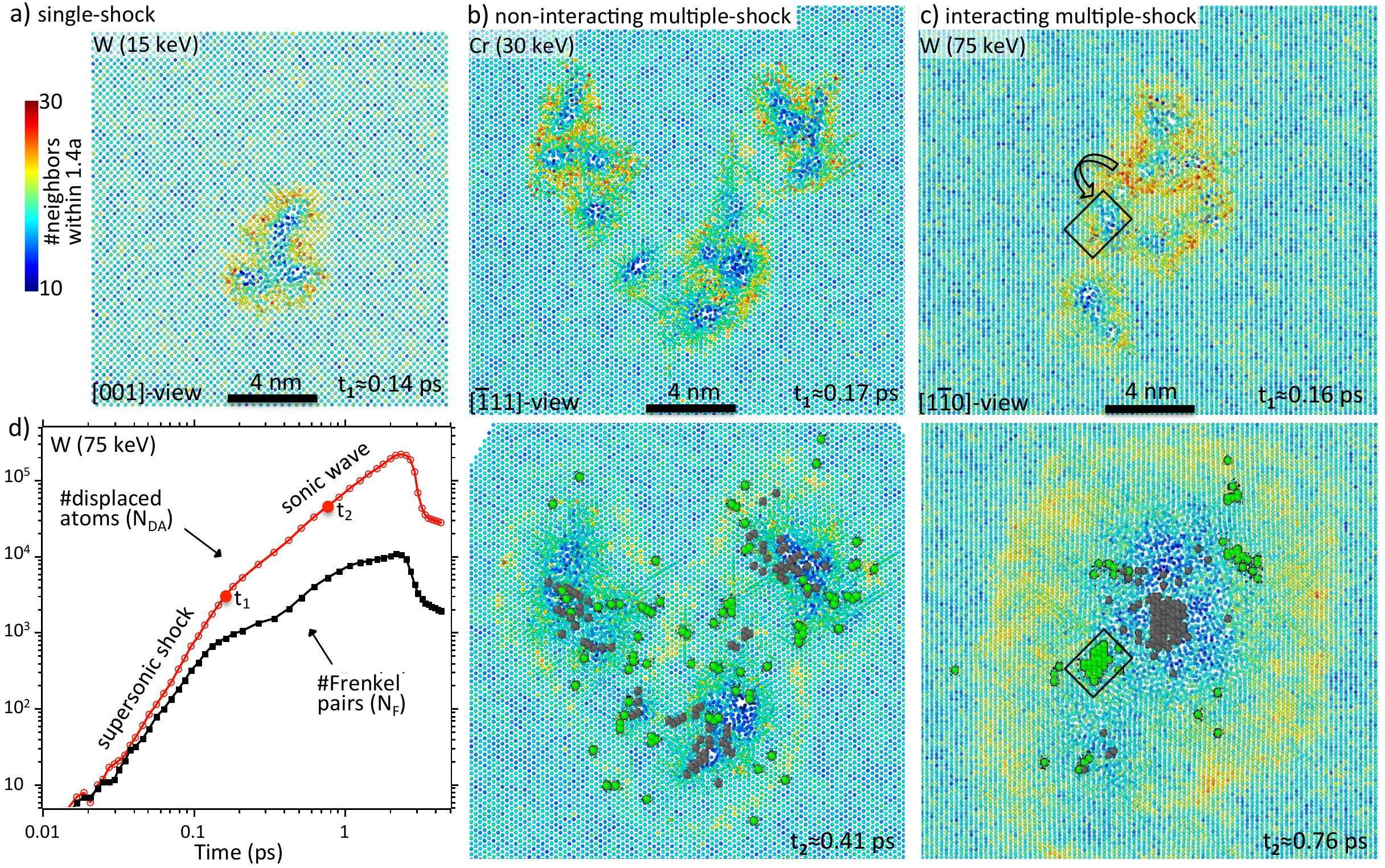}
\caption{(Color online) Colormap of coordination number within $1.4a_0$ revealing three different cascade morphologies a) single-shock (SS), b) non-interacting or separate multiple-shock (sMS) and c) interacting multiple-shock (iMS). Surviving SIAs (green atoms) and vacancies (black) are superimposed to show the associated defect clustering.
\label{fig_map}}
\end{figure*}

To reveal the cascade morphology, a colormap of atom density (number of neighbors within 1.4$a_0$, $a_0$ is the lattice constant) is analyzed as in Ref. \cite{CalderPhilMag10} using the OVITO code \cite{Ovito}. Figure \ref{fig_map}d shows the number of displaced atoms, \NDA, vs. $t$ from a typical high-energy (75 keV) cascade in W. Displaced atoms are those beyond 0.3$a_0$ from any lattice site, and are useful to reveal the pressure wave progression (note that displaced atoms create pressure variation within the materials). At the beginning, up to $\sim$0.18 ps, \NDA\ increases with time as $\sim$$t^3$, suggesting that the wave front moves at a constant speed. For a 75-keV PKA, the speed \cite{supersonic} is ~2.8$\nu_\parallel$ (where $\nu_\parallel$ = 5220 m/s is the speed of a longitudinal sound wave in W). Clearly, the cascade is governed by supersonic shocks at this stage, which ends at $t_1$. The cascade energy is then dissipated elastically via a sonic wave until ~2 ps. The middle of the sonic stage is labeled with $t_2$.

Figure \ref{fig_map}a shows a density map representing the case of \EMD$<$$\mu$. All metals in this regime exhibit a similar morphology consisting of one low-density region (core), even though the core shape is irregular. This figure depicts a single supersonic shock. Hence the cascade exhibits a single-shock (SS) morphology. For brevity throughout this article, the term shock is used to denote a supersonic shock. For \EMD$>$$\mu$, the PKA energy is sufficiently high to create multiple shocks. Figure \ref{fig_map}b shows the morphology associated with Cr at \EMD$>$$\mu$, which is also representative of Fe and Mo. The shocks are separate from each other with a distance similar to the size of the shocks. At $t_2$ (bottom panel), three shock domains are evident. Due to the separation, each shock progresses independently. The surviving SIAs (green atoms) and vacancies (black) are superimposed with the density map (bottom panel) to show how SIAs are distributed at the shock periphery. This morphology is referred to as non-interacting or separate multiple-shock (sMS). Meanwhile, Figure \ref{fig_map}c shows the morphology associated with W at \EMD$>$$\mu$. In this case, the shock fronts are interconnected. The location of the large final SIA cluster is marked by a rectangle. It can be seen that the SIA cluster is located at a low-density region. By following the evolution of such a plot, it is apparent that this cluster originates from the nearby shock fronts as indicated by the arrow. The connectivity among the shocks enables the formation of a high atomic density region and simultaneously provides a place for these atoms to readily form a large cluster. The morphology is referred to as interacting multiple-shock (iMS).
We summarize the cascade morphology classification as:
\begin{itemize}[leftmargin=*]
\item {\it single-shock (SS) morphology}, occurs at \EMD$<$$\mu$. The exponent of the \NF\ curve is $b<1$ due to the energy-dependence of in-cascade SIA-vacancy recombination. 
\item {\it separate multiple-shock (sMS) morphology}, occurs at \EMD$>$$\mu$ e.g. in Cr, Fe and Mo. The high defect survival efficiency in regions in between the shocks balances the in-cascade recombination resulting in $b\approx 1$.
\item {\it interacting multiple-shock (iMS) morphology}, occurs at \EMD$>$$\mu$ e.g. in W. The shocks are closely interconnected allowing the formation of large SIA clusters resulting in $b>1$.
\end{itemize}
Calder \etal\ investigated defect clustering in Fe \cite{CalderPhilMag10}. Using a single Fe PKA, they did not observe interacting multiple shocks. However, when they employed a molecular Bi$_2$ PKA (two Bi atoms), interacting multiple shocks were induced due to high energy deposition associated with the more massive PKA.

Correspondingly, when the cascade morphology is quantified in terms of the presence of supersonic shocks and their connectivity, the characteristic slope of \NF\ curve can be correlated with the underlying cascade morphology. In this sense, the transition energy $\mu$ that marks the beginning of the multiple-shock morphology has a physical importance. While the transition energy falls in a narrow range as previously described, more data points are needed to obtain a more reliable scaling, possibly as a function of mass and/or \Ed. Thus, we have performed additional simulations using the interatomic potentials of Cr and W (representing the lightest and heaviest element in this study, respectively), but applied these potentials to simulation cells that had different atomic mass, as described below (and listed in Table \ref{table_Ed}). As well, we also performed simulations of cascades in tungsten using a different potential labeled as W-bj\"{o}rkas.
\begin{itemize}[leftmargin=*]
\item Cr/W-mass, with Cr potential and W mass for all atoms.
\item Cr/W-PKAmass, with Cr potential and mass for all atoms except the PKA, which was given mass equal to W.
\item W/Cr-mass, with W potential and Cr mass for all atoms.
\item W/Cr-PKAmass, with W potential and mass for all atoms except the PKA, which was given a mass equal to Cr.
\item W-bj\"{o}rkas, simulation using the W potential developed by Bj\"{o}rkas \etal\ \cite{BjorkasNIMB09}.
\end{itemize}

For each of the above systems, the threshold energy has been calculated following the procedure outlined previously. The resulting values of the threshold displacement energy are also provided in Table \ref{table_Ed}. Changing the effective mass of all atoms in the simulation for a given interatomic potential appears to have little to no effect on the threshold energy. However, changing just the mass of the initial PKA atom does result in significant changes, presumably as a result of the scattering efficiency. For W, the threshold energy increases from 122.6 eV to 170.9 eV when the PKA mass is decreased from that of W to that of Cr. Notably, this decrease is reasonably consistent with the decreased scattering efficiency of the energy transfer between Cr and W ($\Lambda  = 4m_1m_2 / (m_1+m_2)^2 \approx$ 0.69). Increasing the PKA mass has an opposite effect. For Cr, \Ed\ decreases from 34.4 eV to 18.9 eV if the PKA mass is changed from Cr to W.

\begin{figure}[htb]
\centering
\includegraphics[width=0.48\textwidth]{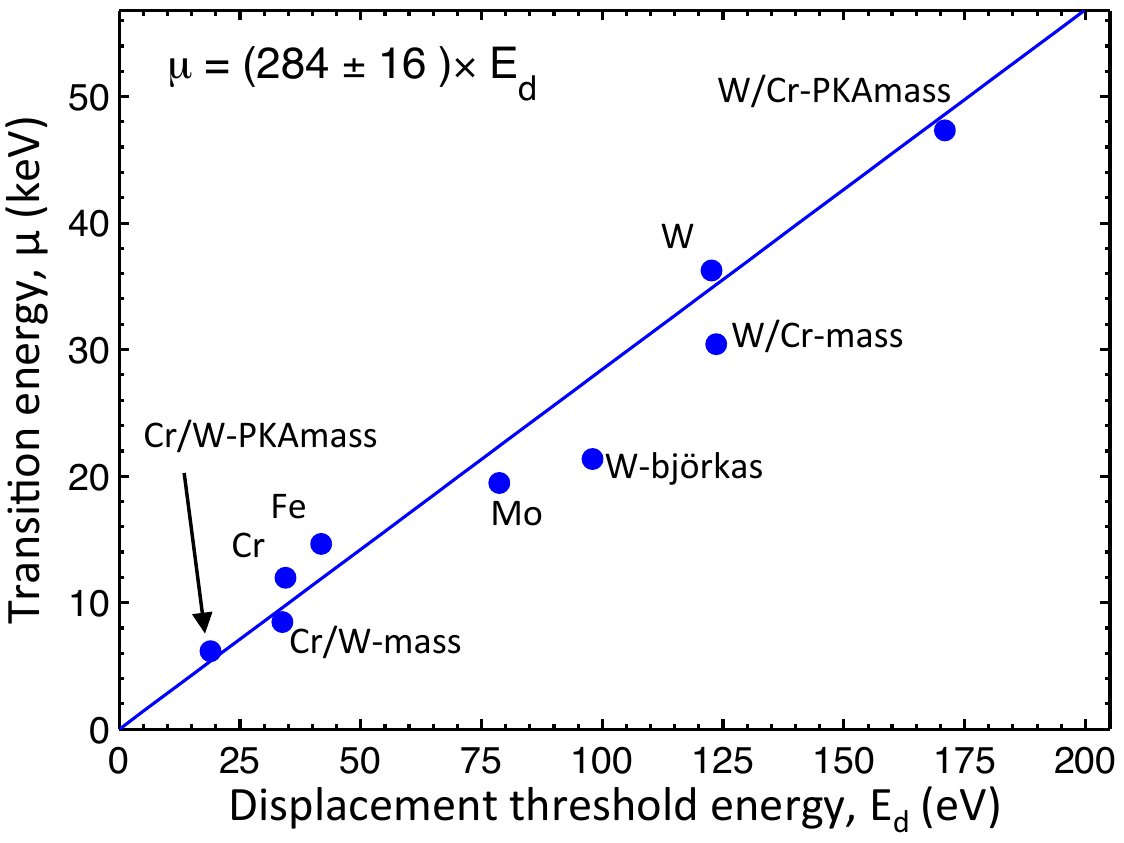}
\caption{Transition energy, $\mu$, as a function of displacement threshold energy, \Ed, showing a good linear scaling.}
\label{fig_transition}
\end{figure}

High-energy cascades were then simulated for these additional systems to determine the transition energy using the default procedure as described previously. Figure \ref{fig_transition} shows the plot of the transition energy as a function of the threshold energy for all systems. Overall, the value of $\mu$ shows a remarkable linear correlation with \Ed. This finding confirms the narrow range of $\mu^*$. A linear fit is performed with a constraint to pass through the origin. The scaling is found to be $\mu = (284 \pm 16) \times E_d$.

The effect of PKA mass on threshold energy seems to be carried over to the transition energy. Increasing the PKA mass, Cr $\rightarrow$ Cr/W-PKAmass, causes the transition to occur at a lower energy. Likewise, decreasing the PKA mass, W $\rightarrow$ W/Cr-PKAmass, increases the transition energy. Despite these shifts, the reduced value of $\mu$ remains unchanged. This indicates that the transition energy is closely related to the threshold energy. For a given PKA energy, increasing the PKA mass causes more damage since \Ed\ is smaller. The probability of forming multiple shocks increases with the damage level. Therefore, the transition energy occurs at a lower value.

When the mass of all atoms is changed altogether, we do not observe a consistent trend in the shift of $\mu$. Increasing the mass, Cr $\rightarrow$ Cr/W-mass, slightly decreases $\mu$. However, decreasing the mass, W $\rightarrow$ W/Cr-mass, also decreases $\mu$. This leads us to conclude that the effect on $\mu$ is insignificant if the mass of all atoms is varied. Note from the threshold simulation that \Ed\ also does not change in such cases. If all atoms still have the same mass, the kinematics of the collisions remains similar. This leads to an insignificant change in \Ed\ and $\mu$. Note that while the cascade kinematics is similar, the dynamics of the cascade evolution occurs faster if the mass is smaller.

Figure 6 shows the \Smax\ for all systems. For the Cr potential, there is a slight increase from Cr to Cr/W-PKAmass. Likewise, for the W potential, there is a slight decrease from W to W/Cr-PKAmass. Above the transition energy, the curves for Fe, Mo, and all cases of Cr still show the behavior characteristic of separate multiple-shock cascade morphology. While those for W-bj\"{o}rkas and all cases of W show the strong clustering associated with the interacting multiple-shock morphology.

\begin{figure}[htb]
\centering
\includegraphics[width=0.48\textwidth]{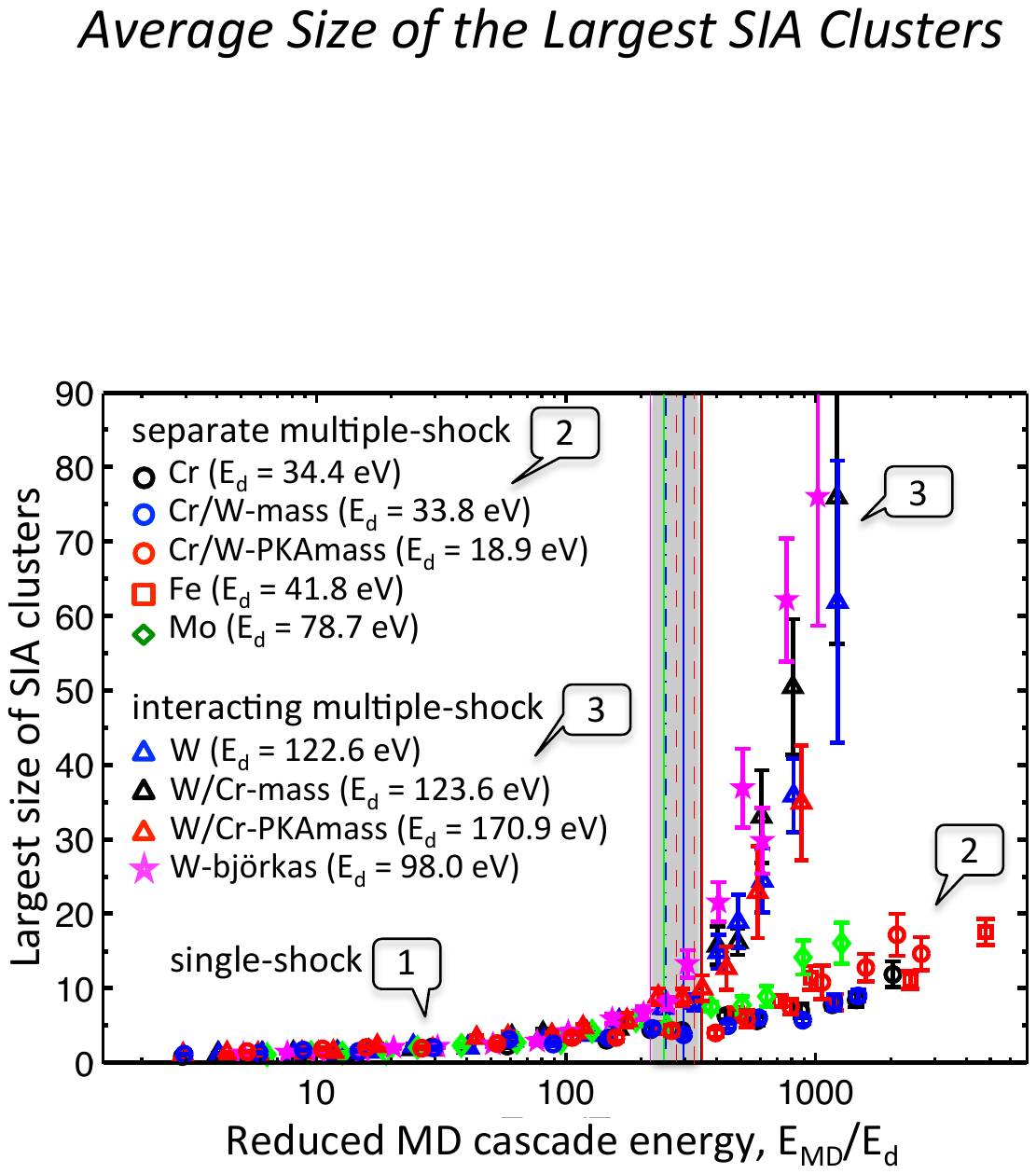}
\caption{(Color online) Average size of the largest SIA clusters vs. \EMD/\Ed\ with standard error. The plots show the three characteristics of SIA clustering associated with the three cascade morphologies. The range of the transition energy is shown as a vertical gray bar.}
\label{fig_maxsia}
\end{figure}

\section{Conclusions}
In conclusion, we have shown that there exists a transition energy in the defect production curve, \NF$\sim$$E_{MD}^b$, of bcc metals. Below the transition energy, $\mu$, only a single supersonic shockwave forms. The value of the exponent is $b<1$. Above the transition, the cascade energy is sufficient to induce multiple supersonic shockwaves. In Cr, Fe, and Mo, the shockwaves are separated from each other. This leads to a value of the exponent, $b \approx 1$. Meanwhile, interconnected shockwaves are observed in W allowing for the formation of large interstitial clusters. In this case the curve exhibits an exponent of $b>1$. The transition energy scales reasonably well with the threshold displacement energy. The empirical fit is found to be $\mu = (284 \pm 16) \times E_d$. All values of \Ed\ have been calculated using a consistent procedure. The procedure takes into account proper weighing factors that include point symmetry and solid angle coverage for each direction grid. For a given potential, the effect of mass on \Ed\ is negligible. However, increasing the PKA mass while keeping the mass of other atoms unchanged decreases \Ed, and vice versa. Due to the linear scaling of $\mu$ with \Ed, the effect of mass on $\mu$ is inherently included in its effect on \Ed. Based on the results of this study, renormalizing the cascade energy with \Ed\ is suggested when comparing defect production among different bcc metals. 

\begin{acknowledgments}
This research has been partially supported by the U.S. Department of Energy, Office of Science, Office of Fusion Energy Sciences (DE-AC06-76RL0-1830), and partially supported by the U.S. Department of Energy, Office of Science, Office of Fusion Energy Sciences and Office of Advanced Scientific Computing Research through the Scientific Discovery through Advanced Computing (SciDAC) project on Plasma-Surface Interactions, under Award No. DE-SC0008875. A part of the computations were performed on Olympus cluster at Pacific Northwest National Laboratory.
\end{acknowledgments}


\begin{thebibliography}{10}
\bibitem{supersonic}
The speed of the wave front, $v_w = \dot{r_e}$, is approximated by equating the number of bcc lattice sites in a sphere of radius $r_e$ to the number of displaced atoms.
\end{thebibliography}
\end{document}